\documentclass[prl,aps,twocolumn,floatfix]{revtex4}

\usepackage{hyperref}
\usepackage{amsmath,amssymb}
\usepackage[utf8]{inputenc}
\usepackage{graphicx,subfigure}
\usepackage{color}
\renewcommand{\vec}[1]{#1}

\newcommand{\dg}{^\dag}
\newcommand{\svk}{_{\vec{k}}}

\newcommand{\num}[1]{#1\dg #1}

\begin{document}
\title{Amplitude-mode dynamics of polariton condensates}
\author{R.~T.~Brierley}
\author{P.~B.~Littlewood}
\affiliation{Cavendish Laboratory, JJ Thomson Avenue, Cambridge, CB3
  0HE, UK} \author{P.~R.~Eastham} \affiliation{School of Physics,
  Trinity College, Dublin 2, Ireland.}

\begin{abstract}
  We study the stability of collective amplitude excitations in
  non-equilibrium polariton condensates. These excitations correspond
  to renormalized upper polaritons and to the collective amplitude
  modes of atomic gases and superconductors. They would be present
  following a quantum quench or could be created directly by resonant
  excitation. We show that uniform amplitude excitations are unstable
  to the production of excitations at finite wavevectors, leading to
  the formation of density-modulated phases. The physical processes
  causing the instabilities can be understood by analogy to optical
  parametric oscillators and the atomic Bose supernova.
\end{abstract}


\maketitle

\newcommand{\tgunkv}[1]{{\vec{#1},t}}
\newcommand{\tgunk}{\tgunkv{k}}

In addition to the phase mode responsible for superconductivity and
superfluidity, Bose-Einstein condensates possess collective amplitude
modes. Examples are the amplitude modes of
superconductors~\cite{littlewood_amplitude_1982} and cold atomic
gases~\cite{altman_oscillating_2002,kollath_quench_2007,barankov_collective_2003,yuzbashyan_relaxation_2006,andreev_nonequilibrium_2004,barankov_atom-molecule_2004,yuzbashyan_integrable_2005},
in which the density fluctuates between condensed and non-condensed
particles, and the formally equivalent Higgs mode of a relativistic
scalar condensate~\cite{varma_higgs_2002}. These modes are orthogonal
to the phase modes in the order-parameter space, and their excitation
is predicted to lead to non-equilibrium states in which the magnitude
of the order parameter oscillates\
\cite{altman_oscillating_2002,kollath_quench_2007,barankov_collective_2003,yuzbashyan_relaxation_2006,andreev_nonequilibrium_2004,barankov_atom-molecule_2004,yuzbashyan_integrable_2005,eastham_quantum_2009,tomadin_signatures_2010}.

The strong coupling of excitons and photons in semiconductor
microcavities leads to the formation of upper and lower
polaritons. Pumping such a microcavity is observed to lead to the
formation of Bose-Einstein condensed states of polaritons,
characterized by coherent light emission from the structure\
\cite{deng_exciton-polariton_2010}. This system allows the
investigation of physics analogous to that of cold atoms, but in a
strongly coupled regime with long range interactions mediated by the
photons. The coupling of polaritons to electromagnetic radiation
outside the microcavity gives an advantage over cold atoms in that
polaritons can be coherently controlled by external pumping, and
conversely their dynamics and coherence can be be directly observed in
the emitted light\
\cite{amo_collective_2010,deng_exciton-polariton_2010}. This has
allowed experiments to reveal the collective behavior arising from
excitations of the phase mode, including
superfluidity~\cite{amo_superfluidity_2009}, vortex
dynamics~\cite{lagoudakis_probing_2011,tosi_onset_2011,sanvitto_persistent_2010},
and the Bogoliubov
spectrum~\cite{utsunomiya_observation_2008,kohnle_single_2011}. However,
the collective behavior associated with~ the amplitude mode has yet to
be considered. Controlled excitation of the amplitude mode of an
atomic Fermi gas requires a rapid switch of the magnetic
field~\cite{barankov_collective_2003,yuzbashyan_relaxation_2006,andreev_nonequilibrium_2004,barankov_atom-molecule_2004,yuzbashyan_integrable_2005},
while in superconductors a coexisting charge-density wave is
needed~\cite{littlewood_amplitude_1982}.

\begin{figure}[t]
  \includegraphics[width=7.5cm]{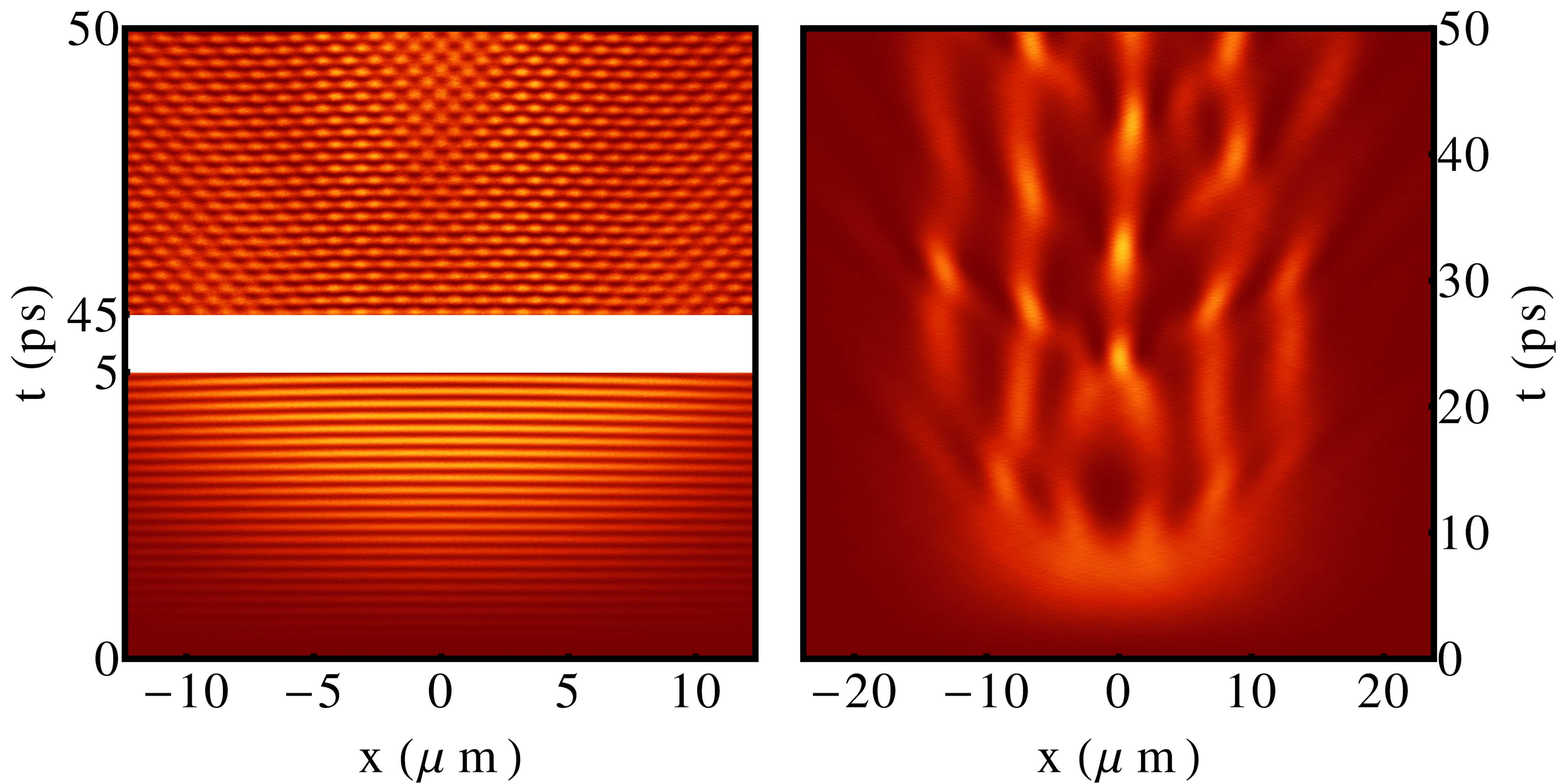}
  \caption{(Color online). Photon density from numerical simulations
    of Eq.  \eqref{eq:gpe}, with resonant pumping of both upper and
    lower polaritons (left panel), and the upper polariton alone
    (right panel). The left panel shows the formation of a
    short-wavelength density modulation due to the wave-mixing
    instability, and the right the result of the attractive-gas
    instability. The pump field profile is a Gaussian at $x=0$ with
    $\sigma=10\, \mu\mathrm{m}$; $\Omega_R=20\,\mathrm{meV}$;
    $E-\omega_0=4\,\mathrm{meV}$.}
  \label{fig:numer-simul-coupl}\vspace{-0.1in}
\end{figure}

The amplitude mode of the polariton condensate may be identified by
considering the Dicke model\
\cite{keeling_bcs-bec_2005,eastham_bose_2001}, which for weak coupling
becomes the BCS model. This limit, with well-known amplitude and phase
modes, is adiabatically connected to the strong-coupling limit
realized in microcavity experiments, where the collective modes are
upper and lower polaritons. Thus the amplitude mode should be
identified with the upper polariton. Uniquely, in this system the
amplitude mode can be directly driven by resonant excitation, allowing
the resulting collective behavior to be studied experimentally. In
this paper we predict the collective behavior arising when the
amplitude degree-of-freedom of a polariton condensate is manipulated
in this way. Whereas driving the phase mode induces superflows, we
find that driving the amplitude mode completely destabilizes the
condensate, causing the polaritons to spontaneously organize into
density-modulated phases (Fig.\ \ref{fig:numer-simul-coupl}). This
occurs because the interactions transfer the excess rest mass and
interaction energy of the non-equilibrium state into kinetic energy,
in a way ruled out for the phase modes by the Landau criterion.

\begin{figure}[tbp]
  \centering
  \includegraphics[width=6.5cm]{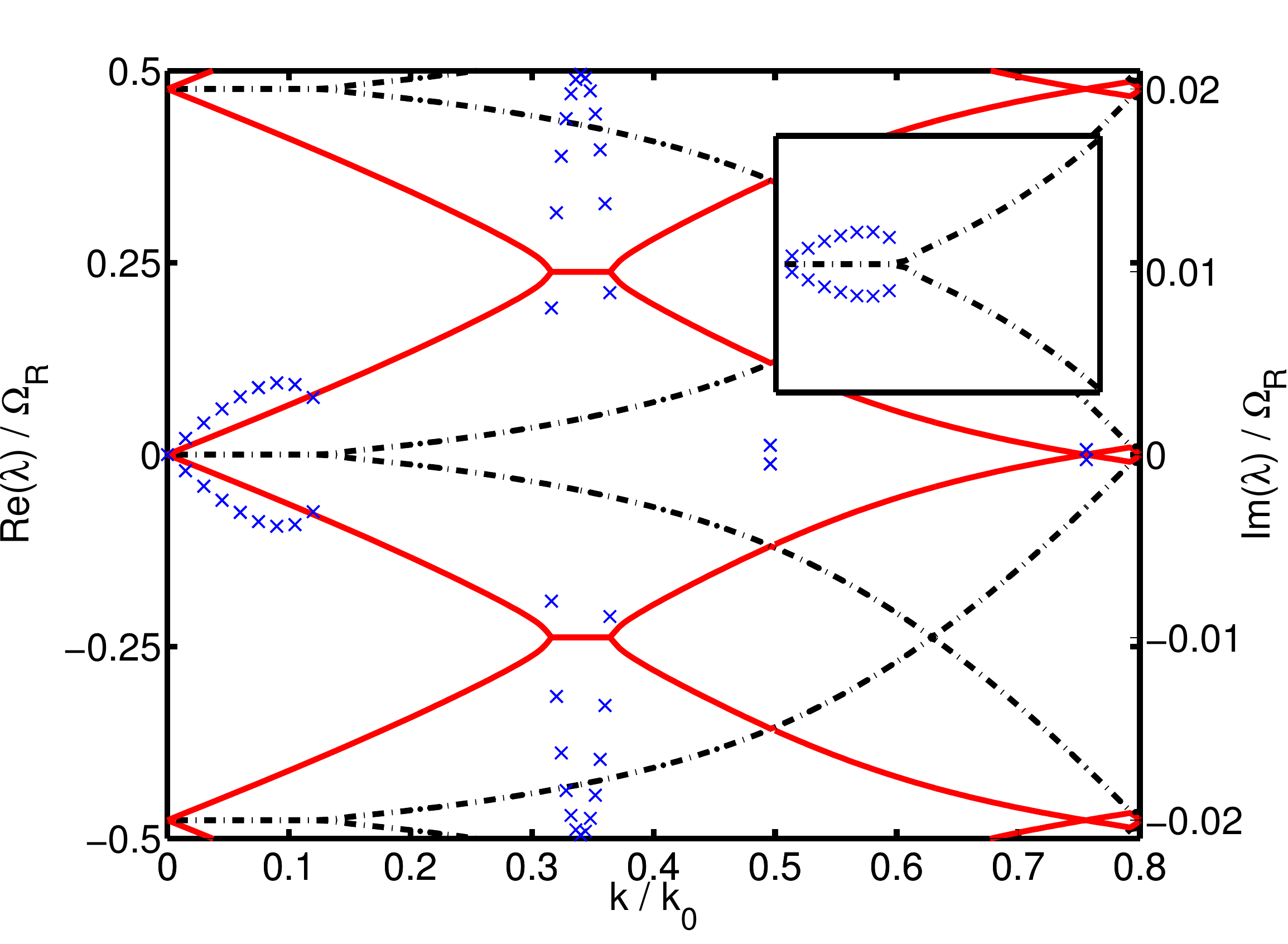}
  \caption{(Color online). Quasi-energy spectrum of an oscillating
    polariton condensate.  Lines (left axis) correspond to the real
    part of the quasi-energy $\lambda$, and blue crosses to the
    non-zero imaginary parts (right axis). Modes with non--zero
    $\Im{\lambda}$ are unstable.  The line colors indicate modes
    deriving from the phase mode/lower polariton (red solid), or the
    amplitude mode/upper polariton (black dashed). Inset: spectrum of
    a weakly-interacting Bose-Einstein condensate with attractive
    interactions, qualitatively reproducing the small-wavevector
    behavior. The parameters are $\omega_0=-E = 0.4 \Omega_R$,
    $\Delta(0) \approx 0.71$, $\phi(0) \approx 0.44$, implying a
    renormalized oscillation frequency $\Omega \approx 0.48\Omega_R$.
  }\label{fig:quasi-energ-lambda}\vspace{-0.1in}
\end{figure}

We begin by considering the dynamics of a Dicke model of polariton
condensation, because for this model exact solutions for the
collective dynamics are
available~\cite{barankov_atom-molecule_2004,andreev_nonequilibrium_2004,yuzbashyan_integrable_2005,babelon_semiclassical_2009}. We
show that these solutions, which describe a uniform oscillating
condensate, are unstable once excitations at finite wavevector are
considered, and systematically identify the instabilities. Our
analysis reveals two types (Fig.~\ref{fig:quasi-energ-lambda}): a
wave-mixing instability between the lower and upper polariton, and a
modulational instability caused by an attractive interaction between
upper polaritons. We propose a Ginzburg-Landau theory which captures
these instabilities, and provides a realistic model of a
microcavity. We use this theory to predict the true steady-states
under continuous excitation.

We can systematically establish the amplitude-mode dynamics of a
polariton condensate by considering the generalized Dicke model\
\cite{keeling_bcs-bec_2005,eastham_bose_2001}
: \begin{equation}\begin{split}
  \label{eq:hamiltonian}
  \hat{H}&=\sum\svk \omega\svk\num{\hat{\psi}\svk}+\sum_i \frac {E}{2}\hat{\sigma}^z_i
  \\ &\qquad+\sum_{i,\vec
    k}\frac{\Omega_R}{2\sqrt{N}}\left(\hat\psi\svk\dg\hat\sigma^-_ie^{-i\vec{k}.\vec{r}_i}+\hat\sigma^+_i\hat\psi\svk
  e^{i\vec{k}.\vec{r}_i}\right)\text{.}\end{split}
\end{equation} This model describes $N$ localized exciton states with
positions $\vec{r}_i$, in the limit where exciton-exciton interactions
exclude double occupancy. Thus each state may be occupied
($\sigma^z_i=1$) or unoccupied ($\sigma^z_i=-1$). The excitons are
coupled to the two-dimensional microcavity photons with in-plane
wavevectors $\vec{k}$, annihilation operators $\hat{\psi}\svk$, and
dispersion relation $\omega_{\vec{k}}\approx
\omega_0+|\vec k|^2/2m_{\text{ph}}$ ($\hbar$=1), where
$m_{\text{ph}}\approx 10^{-5}m_{\text{e}}$.  At this stage we are
neglecting important effects such as the dispersion and inhomogeneous
broadening of the exciton, the polarization degrees-of-freedom~\cite{shelykh_spin_2005,shelykh_pol_2006}, and
the finite lifetime of the cavity photons. We consider only the limit of a single exciton energy $E$, and a single
coupling strength parametrized by the Rabi splitting
$\Omega_R$. To treat the collective dynamics of (\ref{eq:hamiltonian}) it is convenient to work with the expectation values $\psi_\vec{k}=\langle \hat{\psi}\svk\rangle/\sqrt{N}$, $P\svk=\langle \hat{P}\svk\rangle/\sqrt{N}=\frac{1}{N}\sum_i \left<\hat\sigma_i^-\right>
e^{-i\vec{k}\cdot{\vec{r}_i}}$, and $D\svk=\langle \hat{D}\svk\rangle/\sqrt{N}=\frac{1}{N}\sum_i
\left<\hat\sigma_i^z\right> e^{-i\vec{k}\cdot{\vec{r}_i}}$. $\psi\svk$
and $P\svk$ are the macroscopic components of the electric field
and polarization at wavevector $\vec{k}$, while $D\svk$ measures the
exciton occupation.

With only a single photon mode the Heisenberg equations for the model
(\ref{eq:hamiltonian}) are integrable, and closed-form solutions for
the dynamics of the collective variables are
available~\cite{barankov_atom-molecule_2004,andreev_nonequilibrium_2004,yuzbashyan_integrable_2005,babelon_semiclassical_2009}. These
solutions have been studied in the context of quench experiments on
atomic Fermi gases, in which the pairing interaction would be rapidly
switched, giving an impulsive excitation of the collective amplitude
mode (analogous experiments are proposed in light-matter
systems~\cite{eastham_quantum_2009,tomadin_signatures_2010}). This is
predicted to lead to a spatially uniform condensate in which the order
parameter oscillates in time, of which an example is\
\cite{barankov_atom-molecule_2004}
:\begin{eqnarray}\label{eq:zeroformpsi}
  \psi_0&=&\phi(t)=\phi_+\operatorname{dn}(\phi_+ t,\kappa), \\
  P_0&=&\rho(t)=2[-\omega_0\phi(t)+i\dot\phi(t)]/\Omega_R,\label{eq:zeroformp}\end{eqnarray}
where $\mathrm{dn}$ is a Jacobi elliptic function, and the zero of
energy is such that $\psi_0$ is real. The parameters $\phi_+$ and
$\kappa$ determine the period $T$ and magnitude of the oscillations,
and are known functions of the model parameters and the initial
conditions. Note that the oscillation frequency $\Omega=2\pi/T$
generally differs from $\Omega_R$ due to interactions.

Since a microcavity supports a continuum of in-plane modes we must
consider the behavior of an oscillating condensate beyond this
mean-field approximation. To do this we linearize the equations of
motion about the mean-field solution $\psi_0=\phi(t)$, $P_0=\rho(t)$
and $D_0=\Delta(t)$. The fluctuating parts of the collective variables
obey
\begin{gather}
  \label{eq:lineqmom-start}
  i\delta\dot \psi_{\vec k} = \omega_{\vec k}\delta\psi_{\vec
    k}+\Omega_R\delta P_{\vec k}/2\text{,} \nonumber \\
  i\delta\dot P_{\vec k} = E\delta P_{\vec k}-\Omega_R(\Delta
  \delta\psi_{\vec k}+\phi \delta D_{\vec k})/2\text{,} \nonumber \\
  i\delta\dot D_{\vec k} = \Omega_R\left(\rho^*\delta\psi_{\vec
      k}+\phi\delta P^*_{-\vec k}-\phi^*\delta P_{\vec k}-\rho \delta \psi^*_{-\vec k}\right)\text{,}\label{eq:lineqmom-end}
\end{gather}
where the coefficients are time dependent because the condensate
oscillates with angular frequency $\Omega$, which is the gap energy to
the occupied amplitude mode. In deriving (\ref{eq:lineqmom-end}) we
assume that the wavevectors are much smaller than the inverse of the
spacing of the exciton states, so that motional narrowing is effective
and the wavevector of the fluctuations is well defined\
\cite{brierley_finite-momentum_2010,litinskaya_loss_2006}.

Eqs. (\ref{eq:lineqmom-end}) have the form $\dot{\vec
  \Psi}=A(t)\vec\Psi$ where $\vec \Psi$ is a vector of the fluctuating
fields and $A(t)$ a time-periodic matrix. Thus Floquet's theorem
applies and the solutions are of the form $\vec\Psi=\vec
u(t)e^{i\lambda t}$, where $\vec{u}(t+T)=\vec{u}(t)$ and the
quasi-energy $\lambda$ is defined up to an integer multiple of
$\Omega$. Since $u$ is periodic, the stability of the condensate is
determined by the imaginary part of $\lambda$.

To illustrate the fluctuation spectrum of an oscillating condensate,
we show in Fig.~\ref{fig:quasi-energ-lambda} the result obtained for
the exact solution (\ref{eq:zeroformpsi},\ref{eq:zeroformp}). We
measure energy and frequency in units of the Rabi splitting
$\Omega_R$, and wavevector in units of
$k_0=\sqrt{2m_{\mathrm{ph}}\Omega_R}$. The spectrum can be understood
by considering the limit of small oscillations around an equilibrium
condensate, where (\ref{eq:zeroformpsi}) becomes
$\phi=\phi_0+\phi_1\cos\Omega t$ with $\phi_1\ll\phi_0$, and we have
taken the zero of energy to be the equilibrium chemical potential. In
this limit the construction of the spectrum is analogous to that of
the nearly free electron model \cite{ashcroft_solid_1976}, with the
weak time-periodic component playing the role of the weak periodic
potential. To lowest order in the oscillating component, $\phi_1=0$,
the fluctuation spectrum is that of an equilibrium polariton
condensate~\cite{eastham_bose_2001,keeling_bcs-bec_2005}. There is a
mode starting at zero energy and wavevector, which in the low-density
limit is the lower polariton, and in Fig.\
\ref{fig:quasi-energ-lambda} has been renormalized into the
linearly-dispersing phase mode. There is also a gapped mode at zero
wavevector, which in the low-density limit is the upper
polariton~\cite{eastham_bose_2001}, and in
Fig.~\ref{fig:quasi-energ-lambda} is the collective amplitude mode
appearing at the gap frequency $\Omega$
\cite{barankov_synchronization_2006}. The effect of the oscillations
is to fold the spectrum in frequency, and to couple together the
resulting levels. Depending on the phase of the coupling this can
result in either a level repulsion or attraction. In the latter case
the dispersion relation is flattened, and imaginary parts appear for
the quasi-energies. This signals an instability of the spatially
uniform solution, and an initially exponential growth of modes at
finite wavevectors.

The two strongest instabilities in Fig.\ \ref{fig:quasi-energ-lambda}
occur near $|\vec{k}|=|\vec{k}_1|\approx 0.3k_0$ and $|\vec{k}|\approx
0$. The first occurs where the positive-energy branch derived from the
lower polariton crosses with a replica of the corresponding
negative-energy branch. In the low-density limit this occurs only for
$E-\omega_0>0$. The result is a wave-mixing instability in which an
upper and lower polariton from the oscillating condensate at
$\vec{k}=0$ scatter to a pair of lower polaritons at
$\pm\vec{k}_1$. This is analogous to the instability that drives the
microcavity optical parametric oscillator (OPO)\
\cite{ciuti_theory_2003}, with the two $\vec{k}=0$ states forming the
pump, and the states at $\pm \vec{k}_1$ the signal and idler. This
scattering process has previously been considered as a loss mechanism
for incoherent polaritons~\cite{ciuti_branch-entangled_2004}. The
creation of phase modes from amplitude oscillations has been
considered in the Bose--Hubbard model\ \cite{altman_oscillating_2002}
and a similar instability has recently been found in the BCS model\
\cite{dzero_cooper_2009}.

The second instability in Fig.~\ref{fig:quasi-energ-lambda}
corresponds to the flat dispersion relation in the upper polariton
branch near $\vec{k}=0$. It occurs because the saturation of the
light-matter coupling~\cite{rochat_excitonic_2000} reduces the Rabi
splitting with increasing excitation. Thus there is an attractive
interaction between upper polaritons, and states containing more than
a single such excitation are unstable. The resulting form of unstable
spectrum is that obtained from the Bogoliubov analysis for a
condensate with weak attractive interactions (inset).

The microscopic theory is capable of describing the initial
instability, but becomes unwieldy in the nonlinear regime as the
unstable modes evolve. To consider the long-term evolution in a
realistic microcavity we study the Ginzburg-Landau theory
\begin{equation}\begin{aligned}
  i\frac{\partial\psi}{\partial t} &=
  \left(\omega_0-\frac{\hbar^2}{2m_{\text{ph}}}\nabla^2\right)\psi+\frac{\Omega_R}{2}\left(1-\lambda
  |P|^2\right)P 
  \\ & \quad\quad\quad\quad\quad\quad\quad\quad\quad\quad\quad-i \gamma\psi+\xi+F\text{,}\\
  i\frac{\partial P}{\partial t}&=E P+\frac{\Omega_R}{2}(1-\lambda|P|^2)\psi\text{.} \label{eq:gpe}
\end{aligned}\end{equation} Here
$P(x,t)$($\psi[x,t]$) represents the macroscopically-occupied exciton
(photon) field, which is linearly coupled to the photon (exciton)
field to generate polaritons. The nonlinear coupling accounts for the
saturation effect.  This form can be obtained from
(\ref{eq:hamiltonian}) by representing the exciton operators using the
Holstein-Primakoff transformation, and follows directly from the
microcavity exciton-photon Hamiltonian\ \cite{rochat_excitonic_2000}
treated in mean-field theory. We neglect the Coulomb interactions
between the excitons because they act within each quantum well, and
hence are weaker than the saturation nonlinearity when the number of
wells is large. This is consistent with the observed redshifting of
the upper polariton with density in some
microcavities~\cite{kasprzak_bose-einstein_2006}. We have introduced a
damping constant $\gamma=0.25\;\text{ps}^{-1}$ to describe decay of
the cavity photons, a term $F$ to describe a resonant pump laser, and
a noise source $\xi$ to model spontaneous emission noise. We focus on
resonant excitation with circularly-polarized light, and hence include
only a single polarization of exciton and photon. For numerical work
we set $\lambda=1$, absorbing the nonlinear coupling strength into the
definition of density.

Fig.~\ref{fig:numer-simul-coupl} shows the photon densities calculated
from (\ref{eq:gpe}) in one space dimension, with resonant excitation
of both the upper and lower polariton (left panel), and the upper
polariton alone (right panel). For these parameters pumping both modes
initially leads to a uniform induced condensate within the pump spot,
in which as in (\ref{eq:zeroformpsi}) the field amplitude oscillates
due to intermode beating. However, at later times this uniform state
breaks down, and we see the formation of a condensate with a
short-wavelength density modulation. We do not see strong signs of
thermalization and chaos developing from the instability, as has been
suggested for atomic
systems~\cite{dzero_cooper_2009,kollath_quench_2007}, presumably due
to the presence here of dissipation.

In Fig.~\ref{fig:numer-simul-coupl} we also show the behavior when
only the upper polariton is pumped. In this case the wave--mixing
instability cannot occur, and the dominant instability is due to the
attractive interactions. In cold atomic gases, producing a condensate
with attractive interactions leads to the Bose
supernova~\cite{donley_dynamics_2001} where the condensate explodes
due to the excess interaction energy of the uniform state. In the
microcavity, we instead predict the formation of a large wavelength
density modulation. This is because the polaritons can organize in
such a way that the excess energy injected by the pump is
dissipated, leading to a relatively stable steady-state.

While in the undamped model (\ref{eq:hamiltonian}) the uniform state
is always unstable, the dissipation in (\ref{eq:gpe}) implies a
threshold density below which uniform states are stable. This
density $n_0$ is where the gain due to the interactions exceeds the
losses from the modes, and therefore for an interaction strength
$g\sim\Omega_R \lambda$ is $g n_0 \sim \gamma$, up to numerical
factors which are typically of order one. This is essentially the
threshold criterion for the OPO \cite{ciuti_theory_2003}, so that the
threshold density should be within reach experimentally.

The models we have considered can be extended to
include more realistic details. In particular, we have neglected the
electron-hole continuum, which can produce resonant damping of the
upper polariton, similar to the amplitude-mode damping predicted in
atomic systems
\cite{barankov_synchronization_2006,yuzbashyan_dynamical_2006,gurarie_nonequilibrium_2009}. However,
the two-dimensional exciton binding energy in GaAs (CdTe) is $20\,(40)\
\mathrm{meV}$, so that the upper polariton in
Fig.~\ref{fig:numer-simul-coupl} is below the continuum, and not too
strongly damped by this mechanism.

An interesting extension of our work would be to allow both
polarizations of polaritons and excitons in the Ginzburg-Landau
theory. In this case there will be additional amplitude modes
connected to fluctuations in the degree of polarization of the
condensate. It would be interesting to determine whether the
polarization oscillations previously seen in OPO simulations
\cite{shelykh_spin_2005} correspond to these amplitude modes, and to
investigate the possibility of nonlinear decay processes similar to
those discussed here. It would also be useful to investigate the
possibility of attractive interactions between amplitude modes in the
BCS model, and hence establish the extent to which the instabilities
identified here occur in other systems.

In summary, we have used the Dicke model to show that a uniform
condensate in which the amplitude mode is excited is unstable due to
(a)an attractive interaction between amplitude modes and (b)scattering
between amplitude and phase modes. We have used a Ginzburg-Landau
theory to show that these instabilities lead to the formation of
spatially inhomogeneous condensates. In a microcavity the amplitude
mode corresponds to the upper polariton, and therefore these
instabilities can be induced by resonant excitation, leading to features in the real-space density and to bright
emission at an angle from the cavity. Our work shows that there is a
rich collective behavior associated with condensate amplitude modes,
and that microcavities provide a unique opportunity to explore this
physics experimentally.

We acknowledge discussions with J.~Keeling and A.~O.~Silver, and
support from EPSRC-GB and Science Foundation Ireland
(09/SIRG/I1592).

\end{document}